\documentclass[12pt,preprint]{emulateapj}
\usepackage{graphicx}

\shorttitle{Properties of HD~130322}
\shortauthors{Hinkel et al.}

\def\mj{{M$_J$}}
\newcommand{\kms}{km~s$^{-1}$}

\def\gtaprx{ \mathrel{ \vcenter{
      \offinterlineskip \hbox{$>$}
      \kern 0.3ex \hbox{$\sim$}    } } }

\def\ltaprx{ \mathrel{ \vcenter{
      \offinterlineskip \hbox{$<$}
      \kern 0.3ex \hbox{$\sim$}    } } }
      
\def\aj{{AJ}}
\def\apj{{ApJ}}
\def\apjs{{ApJS}}
\def\apjl{{ApJL}}

\def\aap{{A\&A}}
\def\pasp{{PASP}}

\def\mnras{{MNRAS}}

\begin{document} 

\title{Refined Properties of the HD~130322 Planetary System}

\author{
  Natalie R. Hinkel\altaffilmark{1},
  Stephen R. Kane\altaffilmark{1},
  Gregory W. Henry\altaffilmark{2},
  Y. Katherina Feng\altaffilmark{3},
  Tabetha Boyajian\altaffilmark{4},
  Jason Wright\altaffilmark{5,6},
  Debra A. Fischer\altaffilmark{4},  
   Andrew W. Howard\altaffilmark{7}
}
\email{natalie.hinkel@gmail.com}
\altaffiltext{1}{Department of Physics \& Astronomy, San Francisco State University, 
  1600 Holloway Ave, San Francisco, CA, 94132}
\altaffiltext{2}{Center of Excellence in Information Systems, Tennessee
  State University, 3500 John A. Merritt Blvd., Box 9501, Nashville,
  TN 37209}
\altaffiltext{3}{Department of Astronomy \& Astrophysics, 1156 High Street, 
MS: UCO / LICK, University of California, Santa Cruz, CA 95064, USA}
\altaffiltext{4}{Department of Astronomy, Yale University, New Haven,
  CT 06511}
\altaffiltext{5}{Department of Astronomy \& Astrophysics,
  Pennsylvania State University, 525 Davey Laboratory, University
  Park, PA 16802}
\altaffiltext{6}{Center for Exoplanets \& Habitable Worlds,
  Pennsylvania State University, 525 Davey Laboratory, University
  Park, PA 16802}
\altaffiltext{7}{Institute for Astronomy, 2680 Woodlawn Drive,
Honolulu, HI 96822}

\begin{abstract}

Exoplanetary systems closest to the Sun, with the brightest host stars, provide the most favorable opportunities for characterization studies of the host star and their planet(s).  The Transit Ephemeris Refinement and Monitoring Survey uses both new radial velocity measurements and photometry in order to greatly improve planetary orbit uncertainties and the fundamental properties of the star, in this case HD 130322.  The only companion, HD 130322b, orbits in a relatively circular orbit, $e = 0.029$ every $\sim$10.7 days.  Radial velocity measurements from multiple sources, including 12 unpublished from the Keck I telescope, over the course of $\sim$14 years have reduced the uncertainty in the transit midpoint to $\sim$2 hours.  The transit probability for the $b$-companion is 4.7\%, where $M_p \sin i = 1.15 \, M_J$ and $a = 0.0925$ AU.  In this paper, we compile photometric data from the T11 0.8m Automated Photoelectric Telescope at Fairborn Observatory taken over $\sim$14 years, including the constrained transit window, which results in a dispositive null result for both full transit exclusion of HD 130322b to a depth of 0.017 mag and grazing transit exclusion to a depth of $\sim$0.001 mag.  Our analysis of the starspot activity via the photometric data reveals a highly accurate stellar rotation period: 26.53$\pm$0.70 days.  In addition, the brightness of the host with respect to the comparison stars is anti-correlated with the Ca II H and K indices, typical for a young solar-type star. 
\end{abstract}


\section{Introduction}
\label{intro}
Studying the ephemerides, or orbital parameters, of nearby planets is one of the oldest sub-fields in astronomy.  It took nearly 1500 years for a new celestial model to supplant Ptolemy's stationary, geocentric system specifically because, through the use of a number of clever maneuvers (equants, epicycles, and deferents), he was able to accurately predict the motion of the Solar System planets \citep{Gingerich97}.  And it was precisely because Copernicus' heliocentric model did {\it not} predict accurate planetary phenomena that nearly 100 years passed before the theory was generally accepted by the scientific community.  With the purpose of improving orbital uncertainties and fundamental properties of the host star, the Transit Ephemeris Refinement and Monitoring Survey (TERMS) team seeks to characterize individual nearby planetary systems \citep{Kane09}.

The giant planet around HD~130322, a K0 star, was first detected by \citet{Udry00} using the radial velocity (RV) technique via the CORALIE echelle spectrograph.  They reported the period of the planet was $P = 10.720 \pm 0.007$ days, $m\,\sin i = 1.02$ \mj, and eccentricity of $e = 0.044 \pm 0.018$.  The planet was later confirmed by \citet{Butler06}, who observed an additional 12 RV measurements using Keck and determined that the period was $P = 10.70875 \pm 0.00094$ days, $m\,\sin i = 1.089$ \mj, and an eccentricity of $e = 0.025 \pm 0.032$.  The giant planet was further observed by \citet{Wittenmyer09} using both the Hobby-Eberly Telescope (HET) and the Harlan J. Smith telescope.  They found the combination of all four datasets produced a large root-mean-square (rms) variability of 14.8 m\,s$^{-1}$ and an anomalous periodicity at 35 days, specifically due to CORALIE data.  Their orbital parameters, without the RV measurements by \citet{Udry00}, found $P = 10.7085 \pm 0.0003$, $m\,\sin i = 1.04 \pm 0.03 $ \mj, and $e = 0.011 \pm 0.020$. Because of the small eccentricity, \citet{Trilling00} was not able to put a lower bound on the mass of the planet as determined by tidal constraints, only an upper limit of 43.8 \mj.  Observations using the {\it Spitzer} infrared spectrograph by \citet{Dodson11} resulted in the detection of a debris disk around the host star.

Here we present our complete RV data set from a number of sources (including those mentioned above in addition to previously unpublished measurements from the 10m Keck I telescope) that has a time baseline of $\sim$14 years, discussed in \S \ref{star}.  The analysis of the Keplerian orbital solution of these data produce refined orbital ephemeris for the host star HD~130322, with a predicted transit depth of 1.57\%, and 1$\sigma$ transit window of 0.329 days (\S \ref{keplerian}).  In \S \ref{photometry}, we determine the differential magnitude of the host star with respect to multiple comparison stars in order to better understand seasonal and nightly brightness fluctuations.  The evaluation of starspot variability (\S \ref{starspot}) allows us to calculate a stellar rotation period, while understanding the stellar magnetic activity (\S \ref{magact}) gives us insight into the age of HD~130322.

\renewcommand*{\thefootnote}{\fnsymbol{footnote}}

\section{Host Star Properties}
\label{star}
The HD~130322 system has been monitored via the RV technique several times in the past.  We provide a complete RV data set that consists of 118 measurements acquired with CORALIE at the 1.2m Euler-Swiss telescope \citep{Udry00}, 35 measurements acquired with the 2.7m Harlan J. Smith telescope and the High Resolution Spectrograph (HRS) on the HET \citep{Wittenmyer09}, and 24 measurements acquired with the HIRES echelle spectrometer on the 10m Keck I telescope \citep{Butler06}, the most recent 12 of which are previously unpublished.\footnote{Based on observations obtained at the W.\,M.\,Keck Observatory, which is operated jointly by the University of California and the California Institute of Technology.  Keck time has been granted by the University of Hawaii, the University of California, Caltech, and NASA.} We use this combined data set (shown in Table \ref{130322rv}) in order to calculate the fundamental properties of the star as well as the Keplerian 
orbital solution of the planet.

\subsection{Fundamental Parameters}
We derive the host star properties by fitting the high resolution HIRES, HRS, and CORALIE data with Spectroscopy Made Easy \citep{Valenti96}, or SME, via the wavelength intervals, line data, and technique of \citet{Valenti:2005p1491}.  We applied the revised {\it Hipparcos} parallaxes \citep{vanLeeuwen07} to the \citet{Valenti09} methodology as well as surface gravity from a Yonsei-Yale stellar structure model \citep{Demarque04}.  As a result, the fundamental stellar parameters are listed in Table \ref{starparams}, where V-magnitude and distance were determined by {\it Hipparcos}, $B-V$ from Tycho-2,  while effective temperature, surface gravity, projected rotational velocity, stellar mass, and stellar radius were a result of SME.  The high precision of the stellar radius, namely R$_*$ = 0.85$\pm$0.04 R$_{\odot}$, is important when determining the depth and duration of a possible planetary transit.  As a comparison to the SME result, the stellar radius was determined using the Torres relation \citep{Torres10}: R$_*$ = 0.88$\pm$0.04 R$_{\odot}$.  We have also conducted an empirical surface brightness calculation per \citet{Boyajian14} which averages the $V-J$, $V-H$, and $V-K$ surface brightness relations, resulting in an angular diameter of 0.252 $\pm$ 0.006 mas.  Folding in the parallax and error, the radius is 0.89$\pm$0.04 R$_{\odot}$.  All three of these techniques show a very strong consensus for both the stellar radius and error of the host star.  The iron abundance, [Fe/H], as determined by SME, as well as other element abundances, will be discussed in \S \ref{abunds}.  

While our results are consistent with a typical K-type star \citep{Boyajian12}, we note the differences between stellar properties in \citet{Udry00}, namely, their Table 1 and our Table \ref{starparams}.  We have used the updated {\it Hipparcos} \citep{vanLeeuwen07} catalog, which may account for the varying $B-V$ and distance determinations; they cited $B-V =$ 0.781 and the distance to be 29.76 pc.  Our effective temperature, 5387 $\pm$ 44, is also +50 K above their referenced temperatures.

\subsection{Stellar Abundances}\label{abunds}
Stellar abundances have been measured for HD~130322 by at least a dozen different 
groups, for example \citet{Valenti:2005p1491, Neves:2009p1804, DelgadoMena10}.  Due to the proximity of the host star to the Sun, 31.54 pc, a wide variety of elements have been measured within HD~130322, from $\alpha$-type to neutron-capture.  Using the same analysis as seen in {\it Hypatia Catalog} \citep{Hinkel14}, we renormalized the abundance measurements for each dataset to the same solar scale \citep{Lodders:2009p3091} and then determined the maximum measurement variation between the groups, or the {\it spread}, to quantify the consistency of the abundances.   When analyzing the abundances in the {\it Hypatia Catalog}, element measurements were only considered where the {\it spread} was less than the respective error bar associated with that element, or where group-to-group variations were small, and then the median value was used.  For HD~130322, [Fe/H] = 0.12 dex, however the spread between the groups was 0.23 dex.  In other words, the iron ratio was not agreed upon by the various groups, where the renormalized \citet{Ecuvillon:2004p2198} measurement was [Fe/H] = 0.04 and renormalized \citet{Bodaghee:2003p4448} determined [Fe/H] = 0.27 dex.  For this reason, HD~130322 was not included in the analysis (or reduced version) of the {\it Hypatia Catalog}.  Per the SME analysis, [Fe/H] = 0.07 $\pm$ 0.03 dex, using the discussion and solar abundance scale in \citet{Valenti:2005p1491}.  The renormalization gives [Fe/H] = 0.11, which is very close to the median value found for the other data sets in Hypatia.

There were a number of other elements within the star that were measured by different groups.  Per the {\it Hypatia} analysis, where the spread in the abundances were less than respective error and the median value taken: [N/Fe] = -0.14 dex, [Al/Fe] = -0.12 dex, [S/Fe] = -0.14 dex, [Cu/Fe] = -0.15 dex, [Sr/Fe] = 0.07 dex, [YII/Fe] = -0.09 dex, [BaII/Fe] = -0.06 dex, 
[Ce/Fe] = 0.07 dex, [CeII/Fe] = -0.05 dex, and [EuII/Fe] = -0.18 dex.  In general, we find that the majority of abundances well measured in HD~130322 are significantly sub-solar.  Despite measuring [Fe/H] = 0.07 $\pm$ 0.03 via SME, not much can be said conclusively about the overall [Fe/H] content, given the large spread in the abundances determined by different methods.

\section{Keplerian Orbit and Transit Ephemeris}\label{keplerian}
\begin{figure*}
  \begin{center}
    \begin{tabular}{cc}
      \includegraphics[angle=0,width=8.8cm]{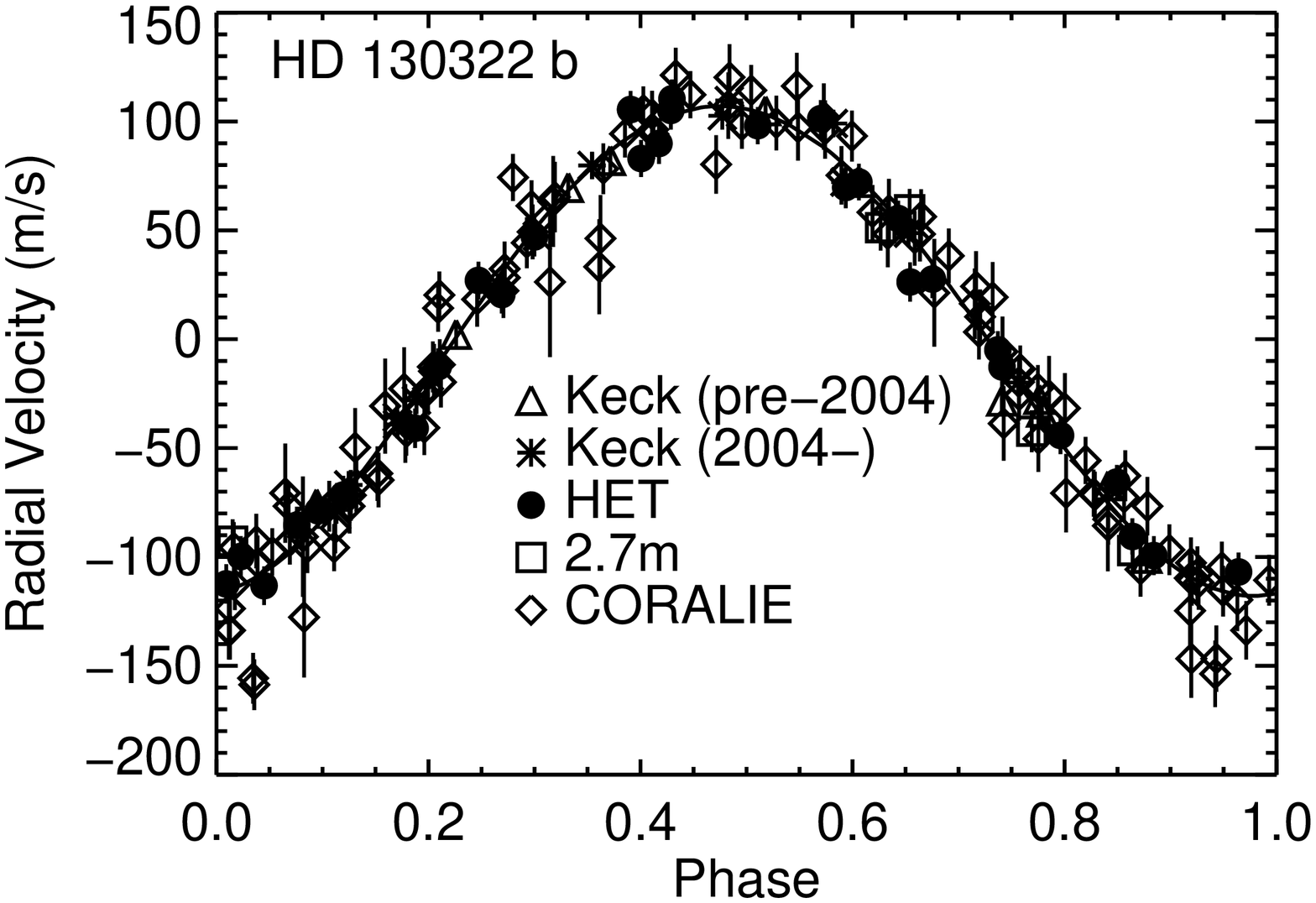} &
      \includegraphics[angle=0,width=8.8cm]{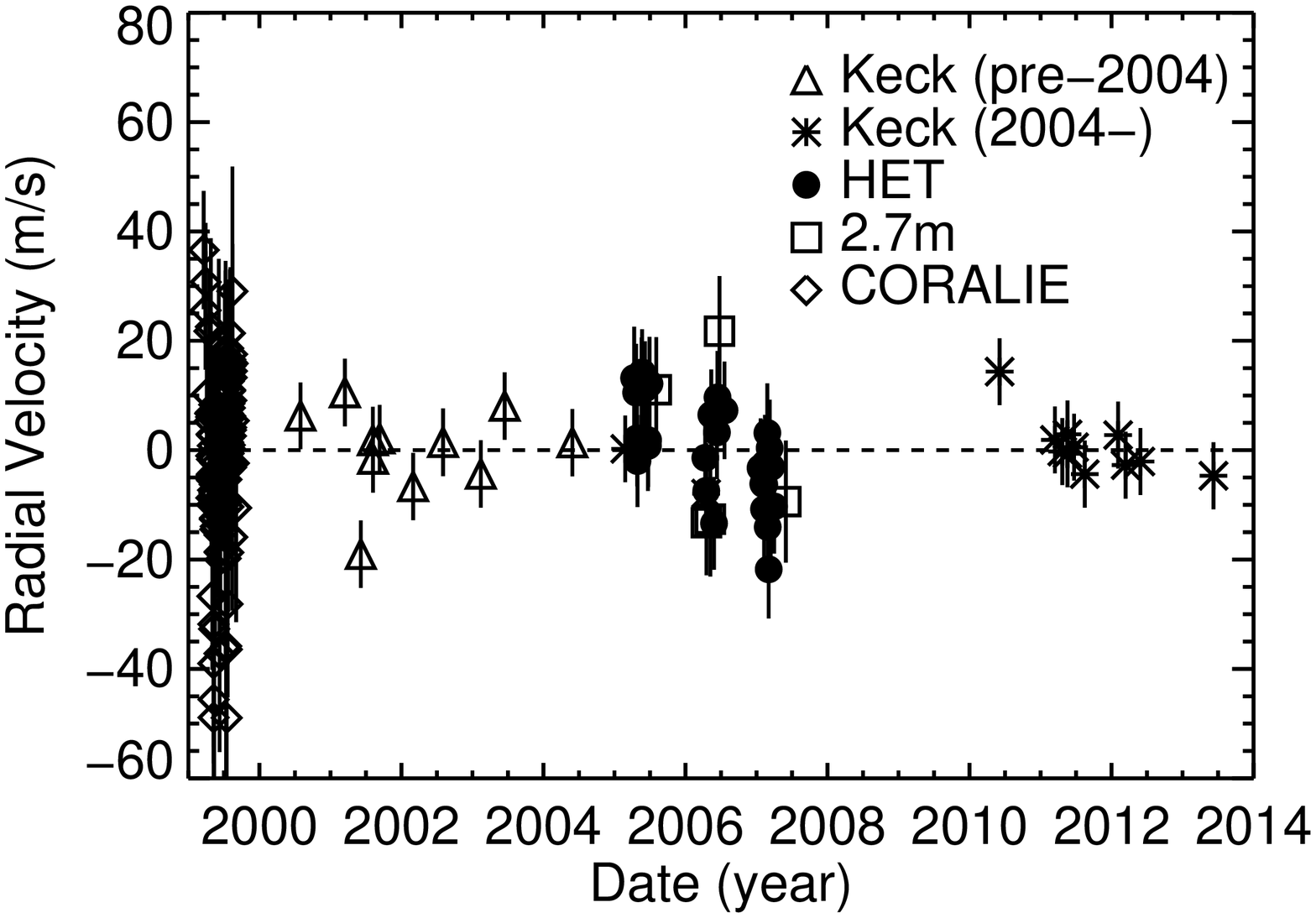}
    \end{tabular}
  \end{center}
  \caption{The Keplerian orbital solution using all of the data shown
    in Table \ref{130322rv}, resulting in the fit parameters shown in
    Table \ref{fitparams}. The typical internal error bars for each data point are plotted. Left: RV data phased on the best-fit
    solution, where the origin of the data is indicated by the different symbols, shown in the figure. Right: Residual velocities with respect to the fitted orbital solution, with the same symbols as the left-side panel.}
  \label{rv}
\end{figure*}

We fit a Keplerian orbital solution to the RV data (shown in Table \ref{130322rv}) using the partially linearized, least-squares fitting procedure described in \citet{Wright09} with parameter uncertainties estimated using the BOOTTRAN bootstrapping routines from \citet{Wang12}. The resulting Keplerian orbital solution is shown in Table \ref{fitparams}, where the stellar parameters for the host star described in Section \ref{star} and summarized in Table \ref{starparams} were used to determine the minimum mass and semi-major axis of the planet. The phased data and residuals to the fit are shown in Fig. \ref{rv}. We find the offsets to be 24.3, 24.7, -27.2, and -23.6 m\,s$^{-1}$ for data from 2.7m McDonald Observatory telescope (2.7m), CORALIE, HIRES (pre-2004 or BJD prior to 13005.5), and HIRES (post-2004), respectively. Regarding the CORALIE data, the median velocity value was subtracted from the data and the velocities were converted from km/s to m/s, which were then used to calculate the offsets with respect to the HET's HRS data.  The fit including the CORALIE data has a $\chi^2_{\mathrm{red}} = 1.35$ and RMS~$= 14.6$~m\,s$^{-1}$. Without the CORALIE data, these numbers change to 1.46 and 8.67, respectively. However, the time baseline of the CORALIE data significantly improves the determination of the orbital period so our fit includes these data for the subsequent analysis.

\LongTables
\begin{deluxetable}{cccc}
\tablecaption{Radial velocities measured for HD 130322 \label{130322rv}}
\tablewidth{0pt}
\tabletypesize{\scriptsize}
\tablehead{
\colhead{BJD} & \colhead{RV} & \colhead{$\pm 1 \sigma$} & \colhead{Tel} \\
\colhead{(--2440000)} & \colhead{(m/s)} & \colhead{(m/s)} & \colhead{}
}
\startdata
11755.76855 & $-$54.8 & 1.0 & HIRES \\
11984.06010 & $-$102.7 & 1.4 & HIRES \\
12065.94426 & $-$55.6 & 1.4 & HIRES \\
12127.80910 & 77.8 & 1.3 & HIRES \\
12128.76352 & 44.8 & 1.4 & HIRES \\
12162.72653 & $-$60.7 & 1.4 & HIRES \\
12335.12452 & $-$127.8 & 1.3 & HIRES \\
12488.77933 & $-$24.9 & 1.5 & HIRES \\
12683.09362 & 54.9 & 1.4 & HIRES \\
12805.91934 & $-$94.5 & 1.4 & HIRES \\
13153.85287 & 42.5 & 1.3 & HIRES \\ --------------- & ----- & ----- & ----------- \\
13426.11560 & $-$43.4 & 1.1 & HIRES \\
13842.00573 & 48.6 & 1.1 & HIRES \\
15351.81672 & 75.5 & 1.1 & HIRES \\
15636.05652 & $-$90.5 & 1.1 & HIRES \\
15673.83685 & 25.2 & 1.2 & HIRES \\
15700.79799 & $-$59.4 & 1.5 & HIRES \\
15700.80123 & $-$62.7 & 1.3 & HIRES \\
15734.89254 & 56.2 & 1.2 & HIRES \\
15789.75142 & 79.1 & 1.3 & HIRES \\
15961.16166 & 86.0 & 1.2 & HIRES \\
16000.02899 & $-$102.0 & 1.2 & HIRES \\
16075.79698 & $-$51.6 & 1.2 & HIRES \\
16451.82279 & 23.7 & 1.2 & HIRES \\
13585.64900 & 83.7 & 7.5 & 2.7m \\
13843.89253 & $-$18.0 & 7.5 & 2.7m \\
13863.78301 & 75.5 & 8.6 & 2.7m \\
13910.78043 & $-$68.5 & 8.1 & 2.7m \\
14251.84318 & $-$72.8 & 9.4 & 2.7m \\
13471.80558 & $-$99.8 & 7.2 & HRS \\
13481.88526 & $-$106.9 & 6.6 & HRS \\
13486.85864 & 105.1 & 6.4 & HRS \\
13488.75815 & 72.2 & 5.9 & HRS \\
13509.79117 & 101.3 & 6.1 & HRS \\
13512.78123 & $-$65.6 & 5.1 & HRS \\
13527.74971 & 27.0 & 6.2 & HRS \\
13542.69985 & 55.4 & 5.6 & HRS \\
13543.70614 & $-$4.9 & 6.1 & HRS \\
13550.70420 & 105.5 & 6.1 & HRS \\
13837.89677 & $-$12.6 & 5.9 & HRS \\
13842.88880 & 27.6 & 6.3 & HRS \\
13868.80896 & $-$78.8 & 5.7 & HRS \\
13882.78043 & 83.0 & 6.0 & HRS \\
13897.72683 & $-$44.2 & 6.1 & HRS \\
13900.72079 & $-$85.4 & 6.0 & HRS \\
13936.63557 & 110.4 & 6.6 & HRS \\
14122.01834 & $-$12.7 & 6.8 & HRS \\
14128.00335 & 47.0 & 6.7 & HRS \\
14135.98084 & $-$113.2 & 6.5 & HRS \\
14139.97029 & 89.9 & 7.2 & HRS \\
14140.96840 & 98.1 & 6.1 & HRS \\                            
14144.96962 & $-$99.4 & 6.6 & HRS \\
14157.01611 & $-$112.4 & 6.8 & HRS \\
14158.92425 & $-$40.9 & 6.8 & HRS \\
14163.92465 & 26.3 & 6.7 & HRS \\
14168.90656 & $-$71.6 & 6.5 & HRS \\
14173.98269 & 69.7 & 7.4 & HRS \\
14176.87914 & $-$90.6 & 5.7 & HRS \\
14191.92631 & 20.5 & 6.2 & HRS \\
11257.85195 & 99.0 & 9.0 & CORALIE \\
11267.80486 & 39.0 & 9.0 & CORALIE \\
11267.81699 & 45.0 & 9.0 & CORALIE \\
11273.86307 & 2.0 & 9.0 & CORALIE \\
11287.71123 & $-$52.0 & 9.0 & CORALIE \\
11287.72337 & $-$68.0 & 9.0 & CORALIE \\
11291.77019 & 137.0 & 9.0 & CORALIE \\
11294.82378 & 44 & 15 & CORALIE \\
11295.76291 & $-$31.0 & 9.0 & CORALIE \\
11296.60937 & $-$72 & 10 & CORALIE \\
11296.83574 & $-$86 & 10 & CORALIE \\
11297.61667 & $-$86 & 10 & CORALIE \\
11297.82760 & $-$99 & 10 & CORALIE \\
11298.60898 & $-$71 & 10 & CORALIE \\
11298.83006 & $-$52 & 10 & CORALIE \\
11299.60924 & $-$18.0 & 9.0 & CORALIE \\
11299.82827 & 2 & 10 & CORALIE \\
11300.60428 & 53.0 & 9.0 & CORALIE \\
11300.82617 & 69 & 10 & CORALIE \\
11301.59933 & 103 & 10 & CORALIE \\
11301.81846 & 119.0 & 9.0 & CORALIE \\
11302.73856 & 105 & 12 & CORALIE \\                          
11304.74462 & 71 & 11 & CORALIE \\
11305.79957 & $-$1 & 10 & CORALIE \\
11306.70099 & $-$58 & 10 & CORALIE \\
11307.78007 & $-$129 & 14 & CORALIE \\
11307.79214 & $-$122 & 14 & CORALIE \\
11308.77008 & $-$131 & 10 & CORALIE \\
11308.78216 & $-$134 & 10 & CORALIE \\
11309.74470 & $-$52 & 11 & CORALIE \\
11309.75677 & $-$47 & 10 & CORALIE \\
11310.65574 & 13 & 10 & CORALIE \\
11310.66781 & 5 & 10 & CORALIE \\
11311.76937 & 51 & 34 & CORALIE \\
11312.71231 & 130.0 & 9.0 & CORALIE \\
11313.70809 & 123.0 & 9.0 & CORALIE \\
11314.71380 & 100 & 12 & CORALIE \\
11315.65132 & 46 & 24 & CORALIE \\
11316.70334 & $-$21 & 14 & CORALIE \\
11317.73465 & $-$81 & 11 & CORALIE \\
11318.71336 & $-$95 & 13 & CORALIE \\
11319.67233 & $-$73.0 & 9.0 & CORALIE \\
11320.73422 & $-$37 & 11 & CORALIE \\                          
11320.74629 & $-$40 & 11 & CORALIE \\
11335.68044 & 141 & 14 & CORALIE \\
11335.69251 & 122 & 14 & CORALIE \\
11336.59905 & 73 & 14 & CORALIE \\
11336.61110 & 84 & 13 & CORALIE \\
11339.65310 & $-$100 & 17 & CORALIE \\
11339.66512 & $-$122 & 17 & CORALIE \\
11340.65091 & $-$109 & 12 & CORALIE \\
11340.66296 & $-$109 & 12 & CORALIE \\
11342.62586 & $-$16 & 11 & CORALIE \\
11342.63788 & 0 & 10 & CORALIE \\
11355.64903 & 119 & 10 & CORALIE \\
11364.57708 & 43 & 11 & CORALIE \\
11366.58262 & 146 & 11 & CORALIE \\
11367.59944 & 124 & 11 & CORALIE \\
11368.57076 & 83 & 11 & CORALIE \\
11369.60470 & 41 & 15 & CORALIE \\
11369.61675 & 49 & 15 & CORALIE \\
11370.51479 & $-$7 & 15 & CORALIE \\
11370.52684 & $-$46 & 17 & CORALIE \\
11373.52546 & $-$66 & 27 & CORALIE \\                          
11373.53762 & $-$103 & 27 & CORALIE \\
11374.54843 & 2 & 18 & CORALIE \\
11374.56049 & $-$14 & 17 & CORALIE \\
11375.55068 & 47 & 11 & CORALIE \\
11375.56276 & 57 & 11 & CORALIE \\
11376.51940 & 58 & 21 & CORALIE \\
11376.53149 & 71 & 19 & CORALIE \\
11380.59301 & 19 & 15 & CORALIE \\
11380.60510 & $-$14 & 16 & CORALIE \\
11381.51359 & $-$46.0 & 9.0 & CORALIE \\
11381.52568 & $-$47.0 & 9.0 & CORALIE \\
11382.48455 & $-$85 & 10 & CORALIE \\
11382.49669 & $-$78 & 10 & CORALIE \\
11382.56236 & $-$83 & 11 & CORALIE \\
11382.57447 & $-$87 & 11 & CORALIE \\
11383.53612 & $-$71 & 10 & CORALIE \\
11383.54681 & $-$88 & 10 & CORALIE \\
11384.54932 & $-$71.0 & 9.0 & CORALIE \\
11384.56142 & $-$62 & 10 & CORALIE \\
11385.54622 & 12 & 10 & CORALIE \\
11385.55834 & 10 & 11 & CORALIE \\                           
11386.54337 & 86 & 10 & CORALIE \\
11386.55545 & 74 & 11 & CORALIE \\
11388.53150 & 132 & 14 & CORALIE \\
11388.54362 & 145 & 14 & CORALIE \\
11389.49681 & 127 & 14 & CORALIE \\
11389.50891 & 121 & 12 & CORALIE \\
11390.46734 & 73 & 11 & CORALIE \\
11390.47944 & 81 & 11 & CORALIE \\
11391.46814 & 5.0 & 9.0 & CORALIE \\
11391.48030 & 11.0 & 9.0 & CORALIE \\
11392.52992 & $-$48.0 & 9.0 & CORALIE \\
11392.54196 & $-$38 & 10 & CORALIE \\
11393.51766 & $-$80 & 10 & CORALIE \\
11393.52971 & $-$91 & 10 & CORALIE \\
11394.47208 & $-$68 & 11 & CORALIE \\
11394.48410 & $-$78 & 11 & CORALIE \\
11395.47053 & $-$25 & 17 & CORALIE \\
11395.48250 & $-$42 & 11 & CORALIE \\
11397.47145 & 88 & 20 & CORALIE \\
11397.48921 & 90 & 15 & CORALIE \\
11398.47062 & 128.0 & 9.0 & CORALIE \\                       
11398.48258 & 121.0 & 9.0 & CORALIE \\
11399.47201 & 139 & 10 & CORALIE \\
11400.48982 & 118 & 10 & CORALIE \\
11401.46724 & 63 & 11 & CORALIE \\
11402.48058 & $-$1 & 17 & CORALIE \\
11403.47387 & $-$52 & 12 & CORALIE \\
11404.47239 & $-$109 & 12 & CORALIE \\
11405.47319 & $-$46 & 22 & CORALIE \\
11406.48267 & $-$6 & 21 & CORALIE \\
11412.48141 & 28 & 11 & CORALIE \\
11412.49345 & 35 & 11 & CORALIE \\
11424.49242 & $-$61 & 20 & CORALIE \\
\enddata
\tablecomments{The line separates the pre-2004 and post-2004 HIRES data. 
}
\end{deluxetable}

\begin{deluxetable}{lcc}
  \tablecaption{\label{starparams} Stellar Parameters}
  \tablehead{
    \colhead{Parameter} &
    \colhead{Value} &
    \colhead{Source}
  }
  \startdata
  $V$                   & $8.04$ & Hipparcos \\
  B-V      & $-0.16$ & Tycho-2  \\
  $Distance$ (pc)      & $31.54 \pm 1.18$  & Hipparcos \\
  $T_{\rm eff}$ (K)                              & $5387 \pm  44$   & SME    \\
  $\log g$                & $4.52 \pm 0.06$   & SME     \\
  $v \sin i$ (km\,s$^{-1}$)          & $0.5 \pm 0.5$  & SME              \\
  M$_*$ ($M_\odot$)                & $0.92 \pm 0.03$    & SME        \\
  R$_*$ ($R_\odot$)                & $0.85 \pm 0.04$  & SME
  \enddata
\end{deluxetable}

\begin{deluxetable}{lc}
  \tablecaption{\label{fitparams} Keplerian Fit Parameters}
  \tablehead{
    \colhead{Parameter} &
    \colhead{Value}
  }
  \startdata
  $P$ (days)                       & $10.70871 \pm 0.00018$ \\
  $T_c\,^{a}$ (JD -- 2440000)      & $16745.594 \pm 0.085$   \\
  $T_p\,^{b}$ (JD -- 2440000)      & $13996.4  \pm 1.1$   \\
  $e$                              & $0.029 \pm  0.016$       \\
  $K$ (m\,s$^{-1}$)                & $112.5 \pm 2.4$        \\
  $\omega$ (deg)                   & $193 \pm 36$        \\
  $\chi^2_{\mathrm{red}}$          & 1.35                   \\
  RMS (m\,s$^{-1}$)                & 14.60
  \enddata
  \tablenotetext{a}{Time of transit.}
  \tablenotetext{b}{Time of periastron passage.}
\end{deluxetable}

The lack of a linear trend over a long period time in Fig. \ref{rv} opens up the possibility to constrain the presence of additional companions in the HD~130322 system.  If $m \sin i$ is the ``minimum mass" (accounting for inclination) then we can also consider the minimum value that $m \sin i$ could possibly have given a linear trend that persists over time (the ``minimum minimum mass"), which is described in detail \citet{Feng15}.  Given that we only have an upper limit on a trend, we have measured the maximum value that the minimum $m \sin i$ could take, or the ``maximum minimum minimum mass" ($Mmmm$).  We use BOOTTRAN to find the 1$\sigma$ maximal value of the linear velocity over $\sim$14 years, where $|dv/dt|$ = 0.0047 m/s/day = 1.716 m/s/yr.  In addition to the values from Table \ref{starparams}, we employ Equation 1 in \citet{Feng15} to calculate $Mmmm$ = 1.83\mj\, as an upper-bound for a possible additional companion in the HD~130322 system.  

Finally, we use the revised orbital properties of the planet described above to derive the predicted transit properties. The predicted time of mid-transit produced by the Keplerian orbital solution is $T_c$=$16745.594 \pm 0.085$ (see Table \ref{fitparams}). Since the orbit is close to circular, the eccentricity has a negligible effect on the transit properties \citep{Kane08}. Using the mass-radius relationship of \citet{Kane:2012p8338}, we adopt a radius of the planet of $R_p = 1.0 R_J$. These combined parameters for the planet result in a transit probability of 4.7\%, a predicted transit duration of 0.16 days, and a predicted transit depth of of 1.57\%. The size of the $1\sigma$ transit window is 0.329 days which can be adequately monitored in a single night of observations \citep{Kane09}.

\section{Photometry}
\label{photometry}
Between Jan 2, 2001 and June 26, 2014, 1569 observations were obtained for HD~130322 at Fairborn Observatory in Arizona using the T11 0.80m APT.  The APT is able to determine the differential brightness of the primary star HD~130322 (P: V = 8.04, B-V = 0.781, K0V) with respect to three comparison stars: HD 130557 (C1: V=6.15, B-V=-0.02, B9V), HD 129755 (C2: V=7.58, B-V=0.41, F2), and HD 132932 (C3: V=7.74, B-V=0.40, F2).  In the initial 2001 observing season, we found that our original comparison star C3 was a low-amplitude variable, so we replaced it with HD~132932 in 2002. Therefore, we have only two comparison stars, C1 and C2, in common for all 14 observing seasons, whereas seasons 2--14 have in common the three comparison stars given above. Like the other telescopes operated on site by Tennessee State University, the Str\"omgren $b$ and $y$ bands are separated and concurrently measured by a photometer with two-channel precision, a dichroic filter, and two EMI 9124QB bi-alkali photomultiplier tubes \citep{Henry99}.  

We compute the six permutations of the differential magnitudes of the four stars in a combinitorical fashion, namely $P-C1$, $P-C2$, $P-C3$, $C3-C1$, $C3-C2$, and $C2-C1$. The magnitudes are then corrected for extinction due to the atmosphere and transformed to the Str\"omgren system, such that the differential $b$ and $y$ observations are combined into a single $(b+y)/2$ band, indicated with the subscript \textit{by}.  To achieve the maximum possible precision, we also combine the three comparison stars to determine the differential magnitudes of HD~130322 with respect to the mean brightness of the comparison stars. The precision of the individual differential magnitudes $P-(C1+C2+C3)/3_{by}$  ranges between $\sim0.0010$~mag and $\sim0.0015$~mag on clearer nights, as determined from the nightly scatter of the comparison stars.  Further details can be found in \citet{Henry99}, \citet{Eaton03}, and references therein.

\begin{figure}
  \includegraphics[width=9cm]{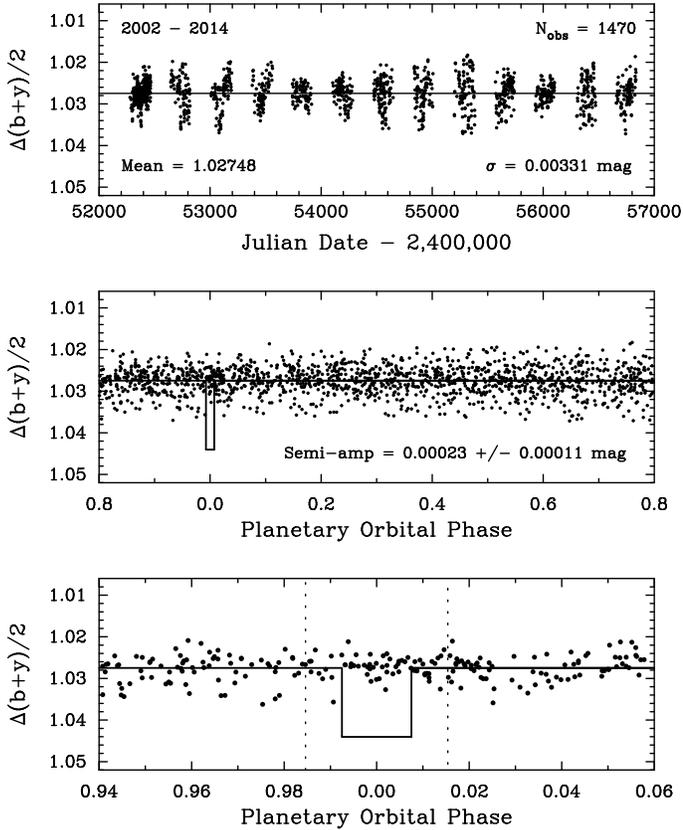}
  \caption{Top: the 1470 differential magnitudes (P - (C1 + C2 + C3) / 3$_{by}$) for HD~130322, taken using the 0.8m APT from 2002-2014, where all 13 observing seasons are normalized to standardize the yearly average. Middle: Observations phased to the planet's ephemeris.  The orbital phase curve semi-amplitude is 0.00023 $\pm$ 0.00011 mag, fit with a least-squares sine, which confirms the b-planet with the lack of periodic light variability in the host star.  Bottom: A zoomed-in portion of the middle-plot, centered on the central transit midpoint.  The solid curve shows the predicted central transit, with a depth of 1.57\% or 0.017 mags and duration of 0.16 days or 0.015 units of phase.  The vertical dotted lines show the $\pm1\sigma$ transit window. Transits are excluded to a depth of 0.017 mag while grazing transits are excluded to a depth of $\sim$0.001 mag.}
  \label{phot}
\end{figure}

The 1470 individual $P-(C1+C2+C3)/3_{by}$ differential magnitudes computed from the 13 observing seasons (2002--2014) are plotted in the top panel of Fig. \ref{phot}. The observations are normalized so that all 13 seasons have the same mean as the first season 2002, indicated by the horizontal line in the top panel, to remove season-to-season variability in HD~130322 caused by a possible starspot cycle (see below). The normalized nightly observations scatter about their grand mean of 1.02748~mag with a standard deviation of $\sigma~=~0.00331$~mag, which is more than a factor of two larger than the $\sim0.0010$ -- $\sim0.0015$~mag measurement precision, which suggests HD~130322 has nightly low-amplitude variation.

The normalized differential magnitudes from the last 13 observing seasons are shown in Fig. \ref{phot} (middle panel), where they are phased with the planetary 10.7 day orbital period and the mid transit time ($T_c$) given in Table \ref{fitparams}.  A fit using a least-squares sinusoid provides a photometric semi-amplitude of $0.00023~\pm~0.00011$~mag and places a one milli-magnitude (0.001~mag) upper bound on the brightness variability of the host star.  In addition, per similar results found in \citet{qhs+2001}, \citet{psch2004}, and \citet{bbs2012}, we dismiss the possibility that jitter induced stellar activity can account for the 10.7-day RV fluctuations.  The constancy of the photometric measurements reveals the true planetary reflex motion seen in the RV variations of HD 130322 is a result of the orbiting planet.

A closer view of the predicted transit window is shown in the bottom panel of Fig. \ref{phot}, plotted with an expanded abscissa. Similar to the middle panel, the solid curves shows the predicted central transit, phased at 0.0, for a duration of 0.16 days or $\sim0.015$ units of phase and a depth of $1.57\%$ or $\sim0.017$~mag.  These values were determined using the stellar radius (Table \ref{starparams}) and orbital ephemeris of the planet (Table \ref{fitparams}).  The $\pm1\sigma$ uncertainty in the transit window timing, as determined by the error bars for both the stellar radius (Table \ref{starparams}) and the improved orbital ephemeris (Table \ref{fitparams}), is indicated by the vertical dotted lines. There are 1405 observations that lie outside the predicted transit window, which have a mean of $1.027490~\pm~0.000089$~mag. The 65 observations that fell within the transit window have a mean of $1.027278~\pm~0.000303$~mag. The difference in these two light levels is our ``observed transit depth," $-0.00021~\pm~0.00032$~mag, consistent to four decimal places. Thus, we are able to rule out full transits, a dispositive null result, with a predicted depth near 0.017~mag and also grazing transits near the predicted time to a depth of $\sim$0.001~mag.

\begin{figure}
  \includegraphics[width=9cm]{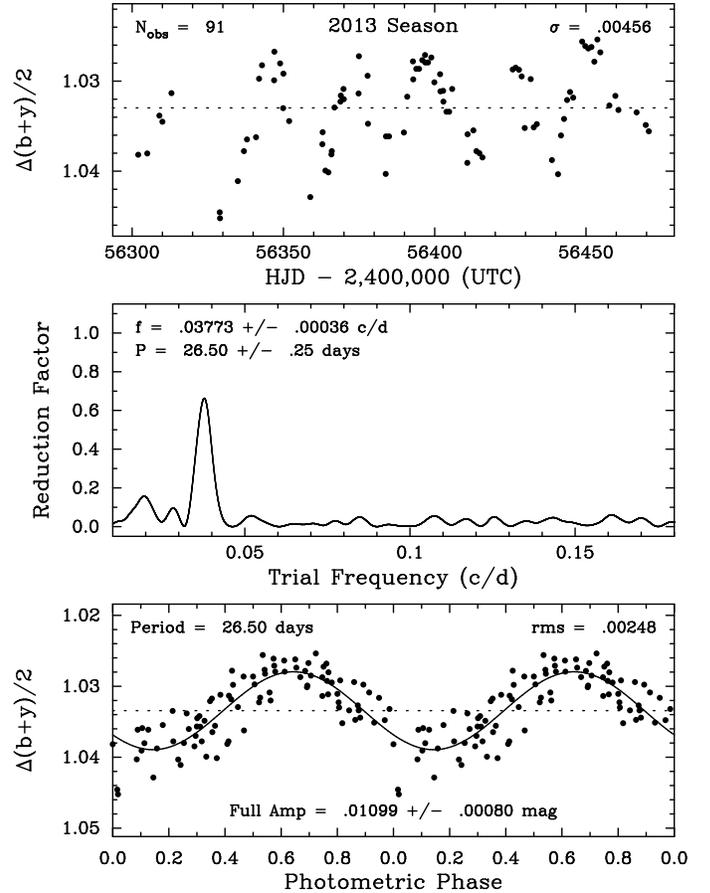}
  \caption{Top: the 91 differential magnitudes (P - (C1 + C2 + C3) / 3$_{by}$) for HD~130322, taken during the 2013 season. Middle: A frequency spectrum of the 2013 observing season of HD 130322, such that the best frequency is at 0.03773 $\pm$ 0.00036 cycles per day (or c/d).  Bottom: the 2013 seasonal observations phased with the corresponding best rotational period of 26.53 $\pm$ 0.70 days. The peak-to-peak amplitude of 0.011 mag shows coherent variability, which may be due to the rotational modulation of starspots, and low noise around the trend. All 14 observing seasons exhibit similar modulation (see Table~\ref{seasons}).}
  \label{season2013}
\end{figure}

\subsection{Starspot Analysis}
\label{starspot}

Given that the scatter of $0.00331$~mag in the normalized data set of Fig. \ref{rv} is significantly larger than the observation precision, we suspect low-amplitude, night-to-night starspot variability in HD~130322.  Inspection of the top panel of Fig. \ref{phot} reveals differences in the amount of scatter from year to year that could also be caused by starspot activity. A solar-type star's rotation period may be determined from the rotational modulation of starspots on the photosphere by measuring the variation in stellar brightness per \citet{Simpson10}. In addition, starspots can resemble an orbiting planet by generating periodic RV fluctuations \citep{qhs+2001}.  Therefore, to determine the behavior of potential starspots, we analyzed all 13 seasons of normalized photometry using a periodogram analysis.  While low-amplitude (0.002--0.017~mag) periodic brightness fluctuations were found during each season, there was no unusual periodicity for the yearly $C2-C1$ comparison stars. The frequency spectrum for the penultimate 2013 observing season is shown in the top panel of Figure \ref{season2013}, while the phase curve is given in the bottom panel.

Results of our complete seasonal period analyses are given in Table \ref{seasons}. Here we include the first observing season from 2001, in which comparison star 3 was later found to be variable and was replaced for the subsequent seasons, as previously mentioned. The period analysis of season 1 (2001) is therefore based on differential magnitudes computed as $P-(C1+C2)/2_{by}$.  Another minor caveat should be noted for the 2004 and 2011 observing seasons.  In both cases, the first harmonic of the rotation period was found to have a slightly higher peak than the rotation period.  For these two seasons, we estimated the rotation periods and their uncertainties by doubling both values.  The mean of the 12 rotation periods, excluding 2004 and 2011, is 26.53~days. The individual rotation periods scatter about that mean with a standard deviation of 2.44~days, and the standard deviation of the mean is 0.70~days.  Therefore, we determine that $26.53~\pm~0.70$~days is our most accurate calculation for the star's rotation period, which matches well with the rotation period derived by \citet{Simpson10}, or $26.1~\pm~3.5$~days, from the first six seasons of our APT data set.  We note the significantly smaller uncertainty in our determination due to the greatly extended 14 year baseline. We found typical starspot-filling factors of one percent or less in Column~6 of Table \ref{seasons}, where the peak-to-peak amplitudes for each season range from 0.002--0.017~mag. Because the stellar rotation period of 26.53~days is more than a factor of two from the 10.7 day planetary RV period and harmonics, in addition to the small impact from starspots, we find that the 10.7-day planet could not be the result of stellar activity.

Examining the stellar rotation with respect to inclination, we first assume that the inclination of rotation axis for HD~130322 is close to $90\arcdeg$, in which case the stellar radius ($0.85~M_{\sun}$) and $v \sin i$ ($0.5$~\kms\,) predict a stellar rotation period of $\sim85.5$~days.  Because this value is over three times longer than our observed value of $P_{rot}=26.53$~days, the logical conclusion is that the stellar rotation axis must have a low inclination and a low planetary transit probability. However, if we substitute the value of $v \sin i = 1.61$ \kms\, from \citet{Butler:2006p6788}, the predicted rotation period is 26.55~days, identical to our observed rotation period within the uncertainties. This implies a very {\it high} inclination near $90\arcdeg$ and, therefore, a high probability of transits. Unfortunately, our photometry unambiguously rules them out.

\subsection{Magnetic Activity}
\label{magact}

\begin{figure}
  \includegraphics[width=8cm]{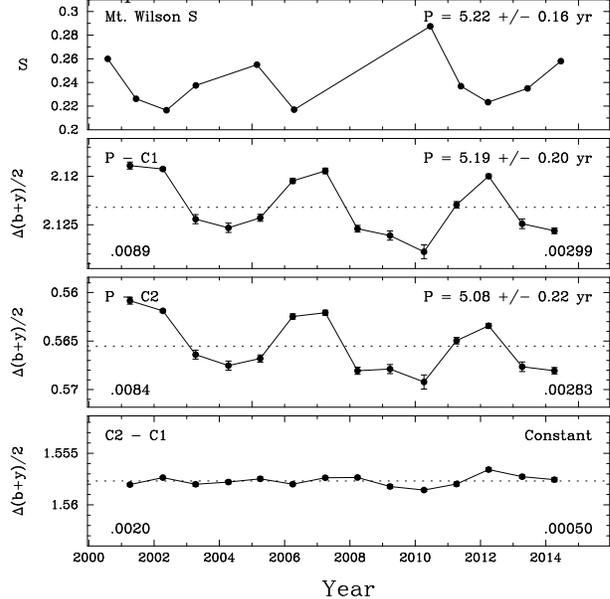}
  \caption{Top: The fluctuations in the Mt. Wilson S-index during the 13 observing seasons. Upper-Middle: Brightness of the primary target with respect to the C1 comparison star, measured with Keck I and the T11 APT. Lower-Middle: Brightness of the primary target with respect to the C2 comparison star. Bottom: Differential magnitudes of the comparison stars, which show stability to 0.0005 mag. The small variability in the two comparison stars, while the primary shows significant fluctuation, and perfect anti-correlation with seasonal mean brightness indicates that the variability in the light curves is intrinsic to HD 130322.  The inverse correlation of S-index and brightness is typical of young, solar-type stars.}
  \label{mags}
\end{figure}

To look for evidence of magnetic cycles in HD 130322, we analyze the variability in the Ca II H and K indices, both proxies for stellar magnetic activity \citep{bal+1995, lsh+2007}, and APT photometry over the entire observing period.  These magnetic cycles could potentially resemble the period of a long-period planet within the RV data.  At the top of Fig. \ref{mags}, we show the seasonal means for the Mount Wilson S-index as determined from the Keck I RV spectra, described in \citet{Wright04,if2010}.  While we do not have the Keck H and K (or RV) measurements for all 13 of our photometric observing seasons, there is notable variability on the order of several years.  The middle two panels show the seasonal mean of HD 130322 (P) varying with respect to the two comparison stars (C1 and C2) throughout the 14-year observations without normalization (Table \ref{seasons}).  The horizontal dotted line indicates the standard deviation of each seasonal mean as compared to the grand mean, given numerically in the lower right corner.  The range in magnitude of the seasonal mean is printed in the lower left corner.  The brightness curves in the middle of Fig. \ref{mags} show HD~130322 varying on the order of multiple mmag with respect to both of the comparison stars.  Similarly, the yearly average of the comparison stars $C2-C1$ is given in the bottom panel, with a very small standard deviation of 0.0005~mag.  Since the comparison stars demonstrate photometric stability over the 13 observing seasons, the fluctuations seen in middle two panels must be intrinsic to the host-star HD~130322.

The variations in H~and~K and APT observations plotted in Fig.~\ref{mags} appear to be cyclic.  Analyses of the yearly means for the Ca II indices, $P-C1$, and $P-C2$ using a least-squares, sine-fit periodogram results in the same periods to within uncertainty: $5.22\pm0.16$, $5.19\pm0.20$, and $5.1208~\pm~0.22$~yr, respectively.  These amplitudes and timescales for HD~130322 are similar to previously monitored  long-term cycles of solar-type stars \citep[see][]{Henry99,lsh+2007,hhl+2009}.

Figure~\ref{mags} reveals that HD~130322's brightness variability is anti-correlated with the strength of its H~and~K emission, as is common among young, lower-main-sequence dwarfs. For example, \citet{lsh+2007} demonstrate the difference in behavior between young, solar-type stars with light curves dominated by dark spots and old solar-type stars with light curves dominated by bright faculae. In the young stars, photometric variability exhibits an inverse (or negative) correlation with chromospheric activity. In older stars, brightness variability and chromospheric activity exhibit a direct (or positive) correlation.  Our Sun exhibits clear direct correlation between total solar irradiance and Ca~II H~and~K emission \citep[see Fig.~2 in ][]{lsh+2007}. \citet{lsh+2007} estimate the dividing line between spot-dominated and faculae-dominated brightness variations to be around $\log{R'_{HK}} = -4.7$.  The original discovery paper of HD~130322b by \citet{Udry00} quoted a $\log{R'_{HK}}$ value of -4.39 from \citet{setal2001}, corresponding to an age of only 0.35~Gyr and a rotation period of approximately 9~days, according to the calibrations in \citep{Wright04}. This predicted rotation period is much shorter than our observed $P_{rot} = 26.53$~days. By way of comparison, \citet{Wright04} demonstrate that the correlation between chromospheric activity and stellar brightness in the G8 dwarf HD~154345 is positive, despite its similar properties to HD~130322.  However, the $\log{R'_{HK}}$ value of HD~154345 is -4.91 compared to -4.78 for HD~130322, which shows it to be much older than HD 130322 so that a positive correlation is expected.

\begin{deluxetable*}{ccccccccc}
\tablenum{4}
\tabletypesize{\small}
\tablewidth{0pt}
\tablecaption{Summary of Photometric Observation for HD 130322}
\tablehead{
\colhead{Observing} & \colhead{} & \colhead{Julian Date Range} &
\colhead{Sigma} & \colhead{$P_{rot}$} & \colhead{Full Amplitude} &
\colhead{$<P-C1>$} & \colhead{$<P-C2>$} & \colhead{$<C2-C1>$} \\
\colhead{Season} & \colhead{$N_{obs}$} & \colhead{(HJD $-$ 2,400,000)} &
\colhead{(mag)} & \colhead{(days)} & \colhead{(mag)} &
\colhead{(mag)} & \colhead{(mag)} & \colhead{(mag)} \\
\colhead{(1)} & \colhead{(2)} & \colhead{(3)} &
\colhead{(4)} & \colhead{(5)} & \colhead{(6)} &
\colhead{(7)} & \colhead{(8)} & \colhead{(9)}
}
\startdata
 2001 &  99 & 51912--52076 & 0.0034 & $23.0\pm0.2$ & $0.0046\pm0.0009$ & $2.1189\pm0.0004$ & $0.5608\pm0.0004$ & $1.5580\pm0.0001$ \\
 2002 & 230 & 52288--52462 & 0.0019 & $24.0\pm0.2$ & $0.0022\pm0.0003$ & $2.1193\pm0.0002$ & $0.5619\pm0.0002$ & $1.5574\pm0.0001$ \\
 2003 &  79 & 52645--52816 & 0.0039 & $29.0\pm0.4$ & $0.0067\pm0.0010$ & $2.1244\pm0.0005$ & $0.5664\pm0.0005$ & $1.5580\pm0.0002$ \\
 2004 &  84 & 53010--53189 & 0.0042 & $31.8\pm0.2$\tablenotemark{a} & $0.0049\pm0.0012$ & $2.1253\pm0.0005$ & $0.5675\pm0.0005$ & $1.5578\pm0.0002$ \\
 2005 &  69 & 53379--53551 & 0.0029 & $29.0\pm0.3$ & $0.0052\pm0.0009$ & $2.1243\pm0.0004$ & $0.5668\pm0.0004$ & $1.5575\pm0.0002$ \\
 2006 &  68 & 53742--53913 & 0.0020 & $24.2\pm0.3$ & $0.0028\pm0.0006$ & $2.1205\pm0.0003$ & $0.5625\pm0.0003$ & $1.5580\pm0.0002$ \\
 2007 &  90 & 54104--54282 & 0.0025 & $24.3\pm0.2$ & $0.0037\pm0.0006$ & $2.1195\pm0.0003$ & $0.5621\pm0.0003$ & $1.5574\pm0.0002$ \\
 2008 &  98 & 54475--54637 & 0.0034 & $28.5\pm0.4$ & $0.0081\pm0.0007$ & $2.1254\pm0.0004$ & $0.5681\pm0.0004$ & $1.5574\pm0.0001$ \\
 2009 &  81 & 54839--55003 & 0.0042 & $30.7\pm0.4$ & $0.0060\pm0.0012$ & $2.1261\pm0.0005$ & $0.5679\pm0.0005$ & $1.5582\pm0.0002$ \\
 2010 &  93 & 55201--55382 & 0.0069 & $26.1\pm0.3$ & $0.0169\pm0.0014$ & $2.1278\pm0.0007$ & $0.5692\pm0.0007$ & $1.5586\pm0.0002$ \\
 2011 &  89 & 55570--55738 & 0.0031 & $24.8\pm0.1$\tablenotemark{a} & $0.0040\pm0.0008$ & $2.1229\pm0.0003$ & $0.5650\pm0.0003$ & $1.5580\pm0.0002$ \\
 2012 &  86 & 55930--56098 & 0.0019 & $25.4\pm0.2$ & $0.0019\pm0.0006$ & $2.1200\pm0.0002$ & $0.5634\pm0.0003$ & $1.5566\pm0.0002$ \\
 2013 &  91 & 56302--56470 & 0.0046 & $26.5\pm0.3$ & $0.0110\pm0.0008$ & $2.1249\pm0.0006$ & $0.5677\pm0.0005$ & $1.5573\pm0.0002$ \\
 2014 &  99 & 56659--56834 & 0.0029 & $27.7\pm0.2$ & $0.0042\pm0.0008$ & $2.1256\pm0.0003$ & $0.5681\pm0.0003$ & $1.5576\pm0.0002$ \\
\enddata
\tablenotetext{a}{Periodogram analysis gave half of the quoted period, 
suggesting that the star had spots on both hemispheres at those epochs.  We 
doubled the photometric periods and their errors in these cases to get
$P_{rot}$.}\label{seasons}
\end{deluxetable*}

\section{Conclusions}
Accurately determining the properties of planetary systems is extremely important as we move towards characterizing the thousands of exoplanets that are now confirmed.  It is only through understanding the host star that we are able to precisely measure the properties of the orbiting planet(s), which fuels both dynamic formation and evolution models.  Through the Transit Ephemeris Refinement and Monitoring Survey (TERMS), we have studied HD~130322 because of the extensive RV coverage offered by HIRES, HRS, and CORALIE over the last $\sim$14 years.  The new and combined data has allowed us to determine a highly precise stellar radius of 0.85$\pm$0.04 $R_{\odot}$, resulting in an updated Keplerian orbital solution to significantly limit the orbital dynamics of the b-planet.  Through an extensive $\sim$14 year photometric baseline at the APT, we carefully monitored the planetary phase during the predicted transit window, which did not reveal any long-term variability of the host-star due to the presence of a companion.  The HD~130322b planet had a transit probability of 4.7\% at a depth of 1.57\%. Significant observations during the predicted transit window yielded a dispositive null result excluding a full transit  to a depth of 0.017 mag and grazing transit to $\sim$0.001 mag.  We were able to quantify the stellar rotation period with an unprecedented accuracy (26.53$\pm$0.70 days) by using the extensive photometric coverage.  The variation in differential magnitudes between the target and reference stars, as compared to the Mt. Wilson S-indices, also allowed us to better understand the stellar magnetic activity.  However, the characterization of the HD~130322 planetary system was only possible through the coming together of both collaborators and techniques, such that we were able to greatly improve the ephemeris of this system.  The TERMS project consistently and systematically provides accurate characterization of bright, nearby planetary systems, forwarding the understanding of exoplanets and their host stars in general.

\section*{Acknowledgements}

The authors would like to thank Howard Isaacson and Geoff Marcy in recognition of their time spent observing the S-indices. N.R.H. would like to acknowledge financial support from the National Science Foundation through grant AST-1109662. The Center for Exoplanets and Habitable Worlds is supported by the Pennsylvania State University, the Eberly College of Science, and the Pennsylvania Space Grant Consortium.  G.W.H. acknowledges long-term support from NASA, NSF, Tennessee State University, and the State of Tennessee through its Centers of Excellence program. T.S.B. acknowledges support provided through NASA grant ADAP12-0172.  J.T.W. and Y.K.F. acknowledge support from the National Science Foundation through grant AST-1211441. A.W.H would like thank the many observers who contributed to the measurements reported here and gratefully acknowledge the efforts and dedication of the Keck Observatory staff. Finally, we extend special thanks to those of Hawai'ian ancestry on whose sacred mountain of Maunakea we are privileged to be guests.  Without their generous hospitality, the Keck observations presented herein would not have been possible.


\begin{thebibliography}{40}
\expandafter\ifx\csname natexlab\endcsname\relax\def\natexlab#1{#1}\fi

\bibitem[{{Baliunas} {et~al.}(1995){Baliunas}, {Donahue}, {Soon}, {Horne},
  {Frazer}, {Woodard-Eklund}, {Bradford}, {Rao}, {Wilson}, {Zhang}, {Bennett},
  {Briggs}, {Carroll}, {Duncan}, {Figueroa}, {Lanning}, {Misch}, {Mueller},
  {Noyes}, {Poppe}, {Porter}, {Robinson}, {Russell}, {Shelton}, {Soyumer},
  {Vaughan}, \& {Whitney}}]{bal+1995}
{Baliunas}, S.~L., {et~al.} 1995, \apj, 438, 269

\bibitem[{Bodaghee {et~al.}(2003)Bodaghee, Santos, Israelian, \&
  Mayor}]{Bodaghee:2003p4448}
Bodaghee, A., Santos, N.~C., Israelian, G., \& Mayor, M. 2003, A{\&}A, 404, 715

\bibitem[{{Boisse} {et~al.}(2012){Boisse}, {Bonfils}, \& {Santos}}]{bbs2012}
{Boisse}, I., {Bonfils}, X., \& {Santos}, N.~C. 2012, \aap, 545, A109

\bibitem[{{Boyajian} {et~al.}(2014){Boyajian}, {van Belle}, \& {von
  Braun}}]{Boyajian14}
{Boyajian}, T.~S., {van Belle}, G., \& {von Braun}, K. 2014, \aj, 147, 47

\bibitem[{{Boyajian} {et~al.}(2012){Boyajian}, {von Braun}, {van Belle},
  {McAlister}, {ten Brummelaar}, {Kane}, {Muirhead}, {Jones}, {White},
  {Schaefer}, {Ciardi}, {Henry}, {L{\'o}pez-Morales}, {Ridgway}, {Gies}, {Jao},
  {Rojas-Ayala}, {Parks}, {Sturmann}, {Sturmann}, {Turner}, {Farrington},
  {Goldfinger}, \& {Berger}}]{Boyajian12}
{Boyajian}, T.~S., {et~al.} 2012, \apj, 757, 112

\bibitem[{{Butler} {et~al.}(2006){Butler}, {Wright}, {Marcy}, {Fischer},
  {Vogt}, {Tinney}, {Jones}, {Carter}, {Johnson}, {McCarthy}, \&
  {Penny}}]{Butler06}
{Butler}, R.~P., {et~al.} 2006, \apj, 646, 505

\bibitem[{Butler {et~al.}(2006)Butler, Wright, Marcy, Fischer, Vogt, Tinney,
  Jones, Carter, Johnson, McCarthy, \& Penny}]{Butler:2006p6788}
Butler, R.~P., {et~al.} 2006, ApJ, 646, 505

\bibitem[{{Delgado Mena} {et~al.}(2010){Delgado Mena}, {Israelian},
  {Gonz{\'a}lez Hern{\'a}ndez}, {Bond}, {Santos}, {Udry}, \&
  {Mayor}}]{DelgadoMena10}
{Delgado Mena}, E., {Israelian}, G., {Gonz{\'a}lez Hern{\'a}ndez}, J.~I.,
  {Bond}, J.~C., {Santos}, N.~C., {Udry}, S., \& {Mayor}, M. 2010, \apj, 725,
  2349

\bibitem[{{Demarque} {et~al.}(2004){Demarque}, {Woo}, {Kim}, \&
  {Yi}}]{Demarque04}
{Demarque}, P., {Woo}, J.-H., {Kim}, Y.-C., \& {Yi}, S.~K. 2004, \apjs, 155,
  667

\bibitem[{{Dodson-Robinson} {et~al.}(2011){Dodson-Robinson}, {Beichman},
  {Carpenter}, \& {Bryden}}]{Dodson11}
{Dodson-Robinson}, S.~E., {Beichman}, C.~A., {Carpenter}, J.~M., \& {Bryden},
  G. 2011, \aj, 141, 11

\bibitem[{{Eaton} {et~al.}(2003){Eaton}, {Henry}, \& {Fekel}}]{Eaton03}
{Eaton}, J.~A., {Henry}, G.~W., \& {Fekel}, F.~C. 2003, in Astrophysics and
  Space Science Library, Vol. 288, Astrophysics and Space Science Library, ed.
  T.~D. {Oswalt}, 189

\bibitem[{Ecuvillon {et~al.}(2004)Ecuvillon, Israelian, Santos, Mayor, Villar,
  \& Bihain}]{Ecuvillon:2004p2198}
Ecuvillon, A., Israelian, G., Santos, N.~C., Mayor, M., Villar, V., \& Bihain,
  G. 2004, A{\&}A, 426, 619

\bibitem[{{Feng} {et~al.}(2015){Feng}, {Wright}, {Nelson}, {Wang}, {Ford},
  {Marcy}, {Isaacson}, \& {Howard}}]{Feng15}
{Feng}, F.~K., {Wright}, J.~T., {Nelson}, B., {Wang}, S., {Ford}, E., {Marcy},
  G.~W., {Isaacson}, H., \& {Howard}, A.~W. 2015, accepted to \apj, arXiv:1501.00633

\bibitem[{{Gingerich}(1997)}]{Gingerich97}
{Gingerich}, O. 1997, {The Eye of Heaven} (Springer)

\bibitem[{{Hall} {et~al.}(2009){Hall}, {Henry}, {Lockwood}, {Skiff}, \&
  {Saar}}]{hhl+2009}
{Hall}, J.~C., {Henry}, G.~W., {Lockwood}, G.~W., {Skiff}, B.~A., \& {Saar},
  S.~H. 2009, \aj, 138, 312

\bibitem[{{Henry}(1999)}]{Henry99}
{Henry}, G.~W. 1999, \pasp, 111, 845

\bibitem[{{Hinkel} {et~al.}(2014){Hinkel}, {Timmes}, {Young}, {Pagano}, \&
  {Turnbull}}]{Hinkel14}
{Hinkel}, N.~R., {Timmes}, F.~X., {Young}, P.~A., {Pagano}, M.~D., \&
  {Turnbull}, M.~C. 2014, \aj, 148, 54

\bibitem[{{Isaacson} \& {Fischer}(2010)}]{if2010}
{Isaacson}, H., \& {Fischer}, D. 2010, \apj, 725, 875

\bibitem[{{Kane} \& {Gelino}(2012)}]{Kane:2012p8338}
{Kane}, S.~R., \& {Gelino}, D.~M. 2012, \pasp, 124, 323

\bibitem[{{Kane} {et~al.}(2009){Kane}, {Mahadevan}, {von Braun}, {Laughlin}, \&
  {Ciardi}}]{Kane09}
{Kane}, S.~R., {Mahadevan}, S., {von Braun}, K., {Laughlin}, G., \& {Ciardi},
  D.~R. 2009, \pasp, 121, 1386

\bibitem[{{Kane} \& {von Braun}(2008)}]{Kane08}
{Kane}, S.~R., \& {von Braun}, K. 2008, \apj, 689, 492

\bibitem[{{Lockwood} {et~al.}(2007){Lockwood}, {Skiff}, {Henry}, {Henry},
  {Radick}, {Baliunas}, {Donahue}, \& {Soon}}]{lsh+2007}
{Lockwood}, G.~W., {Skiff}, B.~A., {Henry}, G.~W., {Henry}, S., {Radick},
  R.~R., {Baliunas}, S.~L., {Donahue}, R.~A., \& {Soon}, W. 2007, \apjs, 171,
  260

\bibitem[{Lodders {et~al.}(2009)Lodders, Plame, \& Gail}]{Lodders:2009p3091}
Lodders, K., Plame, H., \& Gail, H.-P. 2009, Landolt-B{\"o}rnstein - Group VI
  Astronomy and Astrophysics Numerical Data and Functional Relationships in
  Science and Technology Volume 4B: Solar System. Edited by J.E. Tr{\"u}mper,
  4B, 44

\bibitem[{Neves {et~al.}(2009)Neves, Santos, Sousa, Correia, \&
  Israelian}]{Neves:2009p1804}
Neves, V., Santos, N.~C., Sousa, S.~G., Correia, A. C.~M., \& Israelian, G.
  2009, A{\&}A, 497, 563

\bibitem[{{Paulson} {et~al.}(2004){Paulson}, {Saar}, {Cochran}, \&
  {Henry}}]{psch2004}
{Paulson}, D.~B., {Saar}, S.~H., {Cochran}, W.~D., \& {Henry}, G.~W. 2004, \aj,
  127, 1644

\bibitem[{{Queloz} {et~al.}(2001){Queloz}, {Henry}, {Sivan}, {Baliunas},
  {Beuzit}, {Donahue}, {Mayor}, {Naef}, {Perrier}, \& {Udry}}]{qhs+2001}
{Queloz}, D., {et~al.} 2001, \aap, 379, 279

\bibitem[{{Santos} {et~al.}(2001){Santos}, {Mayor}, {Naef}, {Pepe}, {Queloz},
  \& {Udry}}]{setal2001}
{Santos}, N.~C., {Mayor}, M., {Naef}, D., {Pepe}, F., {Queloz}, D., \& {Udry},
  S. 2001, in Astronomical Society of the Pacific Conference Series, Vol. 223,
  11th Cambridge Workshop on Cool Stars, Stellar Systems and the Sun, ed. R.~J.
  {Garcia Lopez}, R.~{Rebolo}, \& M.~R. {Zapaterio Osorio}, 1562

\bibitem[{{Simpson} {et~al.}(2010){Simpson}, {Baliunas}, {Henry}, \&
  {Watson}}]{Simpson10}
{Simpson}, E.~K., {Baliunas}, S.~L., {Henry}, G.~W., \& {Watson}, C.~A. 2010,
  \mnras, 408, 1666

\bibitem[{{Torres} {et~al.}(2010){Torres}, {Andersen}, \&
  {Gim{\'e}nez}}]{Torres10}
{Torres}, G., {Andersen}, J., \& {Gim{\'e}nez}, A. 2010, \aapr, 18, 67

\bibitem[{{Trilling}(2000)}]{Trilling00}
{Trilling}, D.~E. 2000, \apjl, 537, L61

\bibitem[{{Udry} {et~al.}(2000){Udry}, {Mayor}, {Naef}, {Pepe}, {Queloz},
  {Santos}, {Burnet}, {Confino}, \& {Melo}}]{Udry00}
{Udry}, S., {et~al.} 2000, \aap, 356, 590

\bibitem[{Valenti \& Fischer(2005)}]{Valenti:2005p1491}
Valenti, J.~A., \& Fischer, D.~A. 2005, ApJS, 159, 141

\bibitem[{{Valenti} \& {Piskunov}(1996)}]{Valenti96}
{Valenti}, J.~A., \& {Piskunov}, N. 1996, \aaps, 118, 595

\bibitem[{{Valenti} {et~al.}(2009){Valenti}, {Fischer}, {Marcy}, {Johnson},
  {Henry}, {Wright}, {Howard}, {Giguere}, \& {Isaacson}}]{Valenti09}
{Valenti}, J.~A., {et~al.} 2009, \apj, 702, 989

\bibitem[{{van Leeuwen}(2007)}]{vanLeeuwen07}
{van Leeuwen}, F. 2007, \aap, 474, 653

\bibitem[{{Wang} {et~al.}(2012){Wang}, {Wright}, {Cochran}, {Kane}, {Henry},
  {Payne}, {Endl}, {MacQueen}, {Valenti}, {Antoci}, {Dragomir}, {Matthews},
  {Howard}, {Marcy}, {Isaacson}, {Ford}, {Mahadevan}, \& {von Braun}}]{Wang12}
{Wang}, Sharon, X., {et~al.} 2012, \apj, 761, 46

\bibitem[{{Wittenmyer} {et~al.}(2009){Wittenmyer}, {Endl}, {Cochran},
  {Levison}, \& {Henry}}]{Wittenmyer09}
{Wittenmyer}, R.~A., {Endl}, M., {Cochran}, W.~D., {Levison}, H.~F., \&
  {Henry}, G.~W. 2009, \apjs, 182, 97

\bibitem[{{Wright} \& {Howard}(2009)}]{Wright09}
{Wright}, J.~T., \& {Howard}, A.~W. 2009, \apjs, 182, 205

\bibitem[{{Wright} {et~al.}(2004){Wright}, {Marcy}, {Butler}, \&
  {Vogt}}]{Wright04}
{Wright}, J.~T., {Marcy}, G.~W., {Butler}, R.~P., \& {Vogt}, S.~S. 2004, \apjs,
  152, 261

\bibitem[{{Wright} {et~al.}(2008){Wright}, {Marcy}, {Butler}, {Vogt}, {Henry},
  {Isaacson}, \& {Howard}}]{Wright08}
{Wright}, J.~T., {Marcy}, G.~W., {Butler}, R.~P., {Vogt}, S.~S., {Henry},
  G.~W., {Isaacson}, H., \& {Howard}, A.~W. 2008, \apjl, 683, L63

\end{thebibliography}
\end{document}